# Strong ferromagnetic-dielectric coupling in multiferroic $Lu_2CoMnO_6$ single crystals

N. Lee[1], H. Y. Choi[1], Y. J. Jo[2], M. S. Seo[3], S. Y. Park[3], Y. J. Choi[1,a)]

[1]*Department of Physics and IPAP, Yonsei University, Seoul 120-749, Korea*
[2]*Department of Physics, Kyungpook National University, Daegu 702-701, Korea*
[3]*Division of Materials Science, Korea Basic Science Institute, Daejeon 305-806, Korea*

We have first grown single crystals of multiferroic double-perovskite $Lu_2CoMnO_6$ and studied the directional dependence of their magnetic and dielectric properties. The ferromagnetic order emerges below $T_C \approx 48$ K along the crystallographic *c* axis. Dielectric anomaly arises along the *b* axis with no electric polarization at $T_C$, contrary to the polycrystalline work suggesting ferroelectricity along the *c* axis. It is proposed that the incommensurate centric spin modulation leads to the antiferroelectric order with the large dielectric anomaly. Through the strongly coupled ferromagnetic and dielectric states, the highly non-linear variation of both dielectric constant and magnetization was achieved in application of magnetic fields. This concurrent tunability provides a new route to manipulation of multiple order parameters in multiferroics.

[a)]phylove@yonsei.ac.kr

Realization of strong magnetoelectric coupling in multiferroics where ferroelectricity and magnetism coexist, opens new opportunities for novel device applications such as magnetoelectric data storage and sensors utilizing cross-coupling effects between electric and magnetic order parameters.[1-4] The current research on multiferroics is mainly focused on magnetism-driven ferroelectrics in which the ferroelectricity originates from the lattice relaxation via exchange strictions in the ordered magnetic state. Both symmetric and antisymmetric parts of the magnetic exchange coupling can contribute to the ferroelectric distortions. Symmetric exchange interaction is active for multiferroics such as $Ca_3CoMnO_6$ and $GdFeO_3$[5,6] while multiferroicity in spiral magnets of $TbMnO_3$ and $CuBr_2$[7,8] results primarily from antisymmetric exchange interaction. In principle, the substantial coupling between structural distortions and magnetic order can lead to a large variation of the dielectric properties under the application of magnetic fields. However, only few of single phase multiferroics possess the net magnetization, which is advantageous for achieving the mutual control of multiple order parameters.[9,10] In spite of enormous efforts made on multiferroics research, it is still anticipated to design and discover new systems accompanying enhanced cross-coupled functionalities for practical applications. Various systems have been suggested as a candidate belonging to a new class of multiferroics, however, many of them studied so far have been synthesized only in the polycrystalline form, preventing characterization of their intrinsic properties.[11-14]

$Lu_2CoMnO_6$ (LCMO) crystallizes in a monoclinic $P2_1/n$ double-perovskite structure with a unit cell of $a$=0.516 nm, $b$=0.554 nm, and $c$=0.742 nm. $Co^{2+}$ and $Mn^{4+}$ ions are alternatingly located in corner-shared octahedral environments as shown in Figs. 1(a) and (b). It has drawn an interest due to its newly-found multiferroicity in the previous polycrystalline work.[15] The polycrystalline specimen exhibits a broad temperature dependence of dielectric anomaly below ~50 K. It has been predicted that the ferroelectricity along the crystallographic $c$-axis results from the symmetric exchange striction activated by broken inversion symmetry of up-up-down-down (↑ ↑ ↓ ↓) spin arrangement with alternating charge order, similar to the Ising chain magnet of $Ca_3CoMnO_6$. However, this expectation has not yet been confirmed due to the inherent average effect in the polycrystalline form with mixed orientations.

In our single crystals of LCMO, we observed a pronounced peak of dielectric constant along the crystallographic *b*-axis ($\varepsilon'_b$), inconsistent with the previous prediction in the polycrystalline work. We also demonstrate the simultaneous manipulation of interlocked dielectric constant and magnetization by applying magnetic fields. Strong dielectric anomaly with no electric polarization suggests the development of antiferroelectric order arising from the additional incommensurate spin propagation.

We have successfully grown rod-like single crystals of LCMO utilizing the conventional flux method with $Bi_2O_3$ flux. DC magnetization, *M*, was obtained using a SQUID magnetometer (Quantum Design MPMS). Specific heat was measured using the standard relaxation method in a Quantum Design PPMS. Dielectric constant, $\varepsilon'$, was measured in an LCR meter at *f* =10 kHz.

The magnetic properties of LCMO were investigated along the three different crystallographic orientations. The temperature dependence of the magnetic susceptibility in applying magnetic field along the *c* axis, $H_c$ = 3 T was measured and its temperature derivative was taken as shown in Fig. 1(c). As the temperature decreases, the magnetic susceptibility increases smoothly and the ferromagnetic transition occurs at $T_C \approx$ 48 K where the derivative displays the clear anomaly. The sharp peak in the temperature dependence of specific heat divided by temperature also appears at $T_C$ (Fig. 1(d)).

The temperature dependence of magnetic susceptibility, $\chi$, is strongly influenced by the magnitude of applied magnetic fields. As displayed in Fig. 2(a), $\chi$ vs *T* was measured upon warming in *H* = 0.2 T after zero-field cooling (ZFC) and upon cooling in the same *H* (FC). The $\chi$ exhibits the pronounced peak at $T_C$, which can be described as a reentrant spin-glass behavior.[16,17] The temperature at which ZFC and FC curves start to separate relies on the crystallographic orientations. The isothermal *M* at 5 K is shown in Fig. 2(b). The large ferromagnetic hysteresis in $M_c$ was observed whereas $M_a$ and $M_b$ show almost linear field dependences with weak hystereses. $M_c$ is saturated at about 3 T with a magnetic moment of ~6 $\mu_B$, consistent with the summation of $Co^{2+}$ (*S* = 3/2) and $Mn^{4+}$ (*S* = 3/2) magnetic moments in a formula unit. The saturation magnetic field in single crystalline specimen of ~ 3 T is much smaller than the value of ~ 60 T at 0.5 K in the polycrystalline one. This indicates that the single crystal does not implicate a significant pinning effect that is ordinarily present in a polycrystalline form due to grain boundaries and defects.

The temperature dependence of dielectric constants along the three different crystallographic orientations in zero magnetic field is presented in Fig. 3(a). Surprisingly, the broad peak with clear anomaly at $T_C$ is observed only in $\varepsilon'_b$ whereas both $\varepsilon'_a$ and $\varepsilon'_c$ show no distinct anomalies, contrary to the predicted $c$-axis ferroelectricity in the polycrystalline work. However, no measurable electric polarization is detected in the pyroelectric current measurement which is highly sensitive to a polarization magnitude. We note that the variation of $\varepsilon'_b$ below $T_C$ is much enhanced compared to that in the polycrystalline sample. In our LCMO crystals, the peak height in $\varepsilon'_b$ normalized by the value at $T_C$ appears to be ~15 %, but the variation in polycrystalline sample is only ~2 %.

The evident dielectric constant anomaly along the $b$-axis without any observable electric polarization can be interpreted as the antiferroelectric-type scenario. The E-type antiferromagnet of orthorhombic perovskite HoMnO$_3$ is known as a prototype multiferroic where the ferroelectricity along the crystallographic $a$ axis is driven by Mn-Mn symmetric exchange striction.[18] However, in a single crystalline form, only the dielectric anomaly occurs along the $a$ axis.[19] It turns out that the centric spin density wave with the incommensurate E-type antiferromagnetic order is realized and the inversion center coinciding with the atomic structure leads to the antiferroelectric order. In LCMO, the $ab$ plane spin modulation is found to be very slow and incommensurate as $(k_a, k_b) \approx (0.0223, 0.0098)$.[15] Similar to orthorhombic HoMnO$_3$, the long-wavelength spin-density wave may be responsible for no observable electric polarization. Therefore, further investigations are desirable in order to establish the exact spin configurations projected in the plane and also to reveal the mechanism for the spin-driven antiferroelectricity.

Fig. 3(b) exhibits the temperature dependence of $\varepsilon'_b$ in various magnetic fields, $H_c$ = 0, 1, 1.5, 1.8, 2, and 3 T. The broad peak in $\varepsilon'_b$ is gradually suppressed as $H_c$ increases and it completely disappears at $H_c$ =3 T where the magnetization is saturated. The most discernible change occurs between 1 and 2 T in accordance with the precipitous rise in isothermal magnetization in Fig. 2(b), suggesting that ferromagnetic and dielectric properties are strongly interconnected. This reduction of the peak in $\varepsilon'_b$ can be explained by the destruction of the $ab$ plane spin propagation upon increasing $H_c$ with development of the fully saturated ferromagnetic moment.

Taking advantage of the delicate response of $\varepsilon'_b$ to the external magnetic fields, the magnetic control of $\varepsilon'_b$ interlocked with $M_c$ in a wide range of the temperature is attained as shown in Fig. 4. The variation of $H_c$ up to 4 T gives rise to both a highly non-linear decrease of $\varepsilon'_b$ on the order of maximum 20 % (Fig. 4(a)), and an increase of $M_c$ (Fig. 4(b)). The most pronounced variation occurs at 35 K, the peak position of $\varepsilon'_b$ (*T*) (Fig. 3). At 5 K, $\varepsilon'_b$ and $M_c$ show the largest magnetic hysteresis. Upon increasing the temperature, the hysteretic behavior diminishes and almost disappears at 45 K near the magnetic transition temperature. Note that the coupling between dielectric and ferromagnetic order parameters is very rare[20] since spin configuration in most of magnetism-driven ferroelectrics is based on frustrated antiferromagnetic orders. This magnetodielectric effect in a ferromagnet provides practical benefit for application in magnetic device.

In summary, we have successfully synthesized single crystals of new multiferroic $Lu_2CoMnO_6$ and explored magnetic and dielectric properties in different crystallographic orientations. A large and broad peak of dielectric constant along the *b*-axis with no electric polarization was observed, supposedly described by the antiferroelectric order from centric incommensurability of spins. Furthermore, the magnetic control of dielectric constant was accomplished by the intimate correlation between ferromagnetic and dielectric states. Our findings offer important clues for understanding microscopic mechanism for multiferroicity of this compound and also present capability of a practical memory storage utilizing both magnetic and dielectric quantities.

**Figure Captions**

**FIG. 1** (Color online). (a) & (b) Views of the crystal structure of double perovskite $Lu_2CoMnO_3$ ($P2_1/n$) from the $a$ axis (a) and from the $c$ axis (b). Orange, violet, light blue and yellow spheres represent $Lu^{3+}$, $Co^{2+}$, $Mn^{4+}$, and $O^{2-}$ ions, respectively. The grey box with the cross-section rectangles designates the crystallographic unit cell. (c) Temperature dependence of magnetic susceptibility, $\chi=M/H$, (1 emu=$4\pi \times 10^{-6}$ m$^3$) in $H_c$=3 T and its temperature derivative up to 150 K. (d) Temperature dependence of specific heat divided by the temperature, $C/T$. Dotted line indicates the Curie temperature of ~48 K.

**FIG. 2** (Color online). (a) Temperature dependence of $\chi$ measured up to 300 K upon warming in $H$=0.2 T after zero-field cooling (ZFC) and upon cooling in $H$=0.2 T (FC) along the $a$, $b$, and $c$ axes. (b) Isothermal magnetization, $M$, with both ramping up and down measurements up to 7 T along the $a$, $b$, and $c$ axes at 5 K.

**FIG. 3** (Color online). (a) Temperature dependence of dielectric constant, $\varepsilon'$, below 80 K along the $a$, $b$, and $c$ axes in zero magnetic field. (b) Temperature dependence of dielectric constant along the $b$ axis, $\varepsilon'_b$, below 80 K in $H_c$=0, 1, 1.5, 1.8, 2, and 3 T.

**FIG. 4** (Color online). (a) Magnetic field dependence of $\varepsilon'_b$ with ramping up and down $H_c$ up to 4 T in various temperatures, $T$=5, 10, 20, 35, and 45 K. (b) Isothermal magnetization along the $c$ axis, $M_c$, up to 4 T at $T$=5, 10, 20, 35, and 45 K.

**Acknowledgements**

We thank H. Eisaki and Y. Yoshida at AIST for their assistance with the magnetization measurement. Work at Yonsei University was supported by the NRF Grant (NRF-2012M2B2A4029730 and NRF-2013R1A1A2058155) and BK21 Plus project. Work at Kyungpook National University was supported by the NRF grant (NRF-2013R1A1A2063904). Work at KBSI was supported by the Pioneer Research Center Program through the National Research Foundation of Korea funded by the Ministry of Science, ICT & Future Planning (2011-0027908).

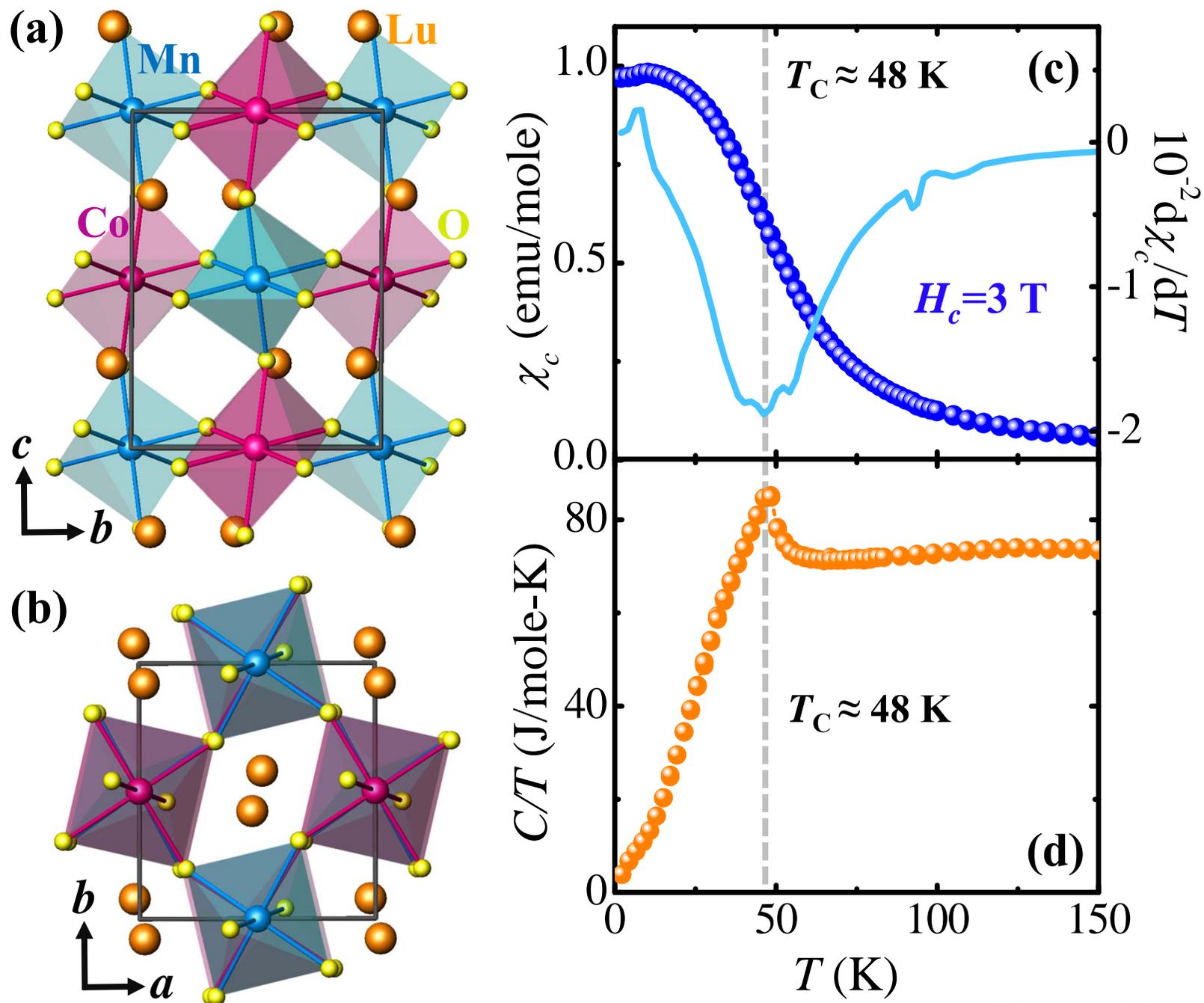

Fig. 1 N. Lee et al.

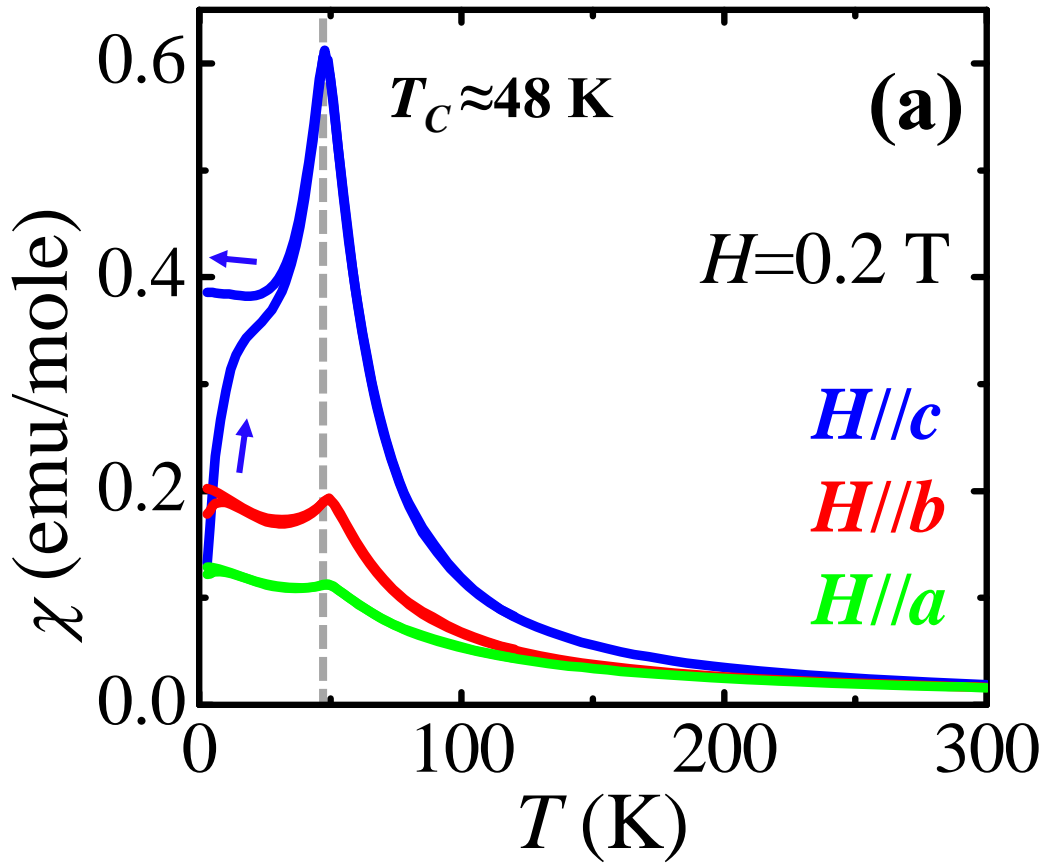
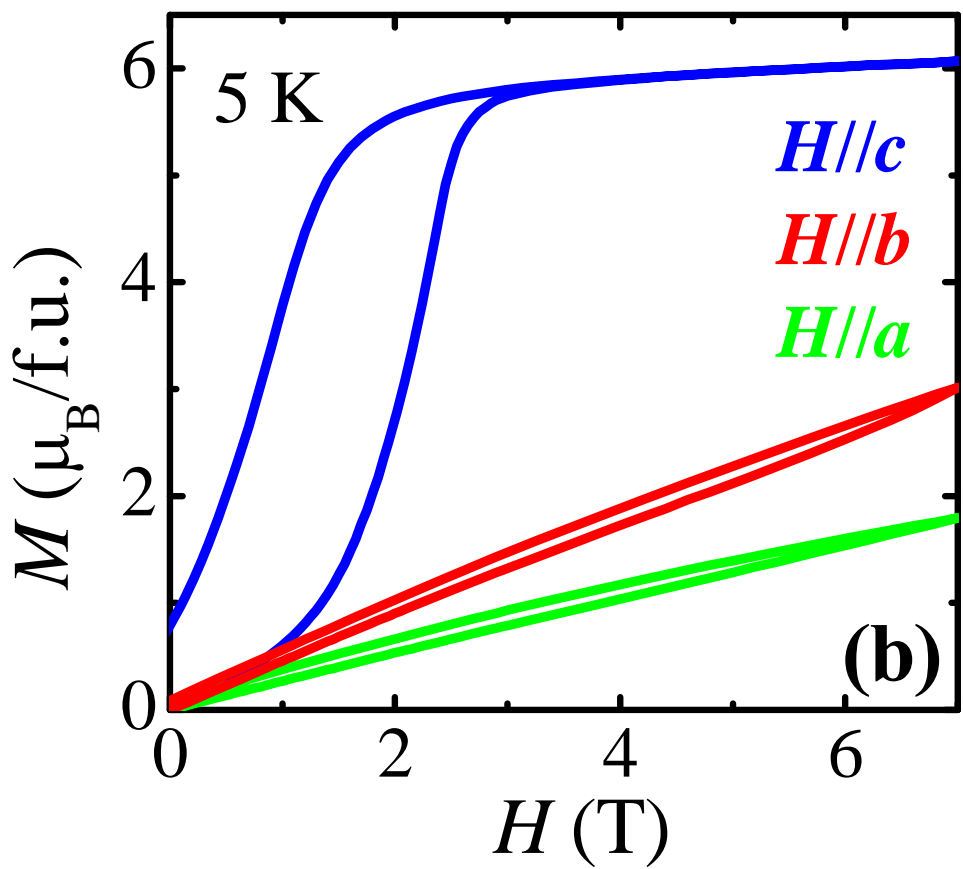

Fig. 2 N. Lee *et al.*

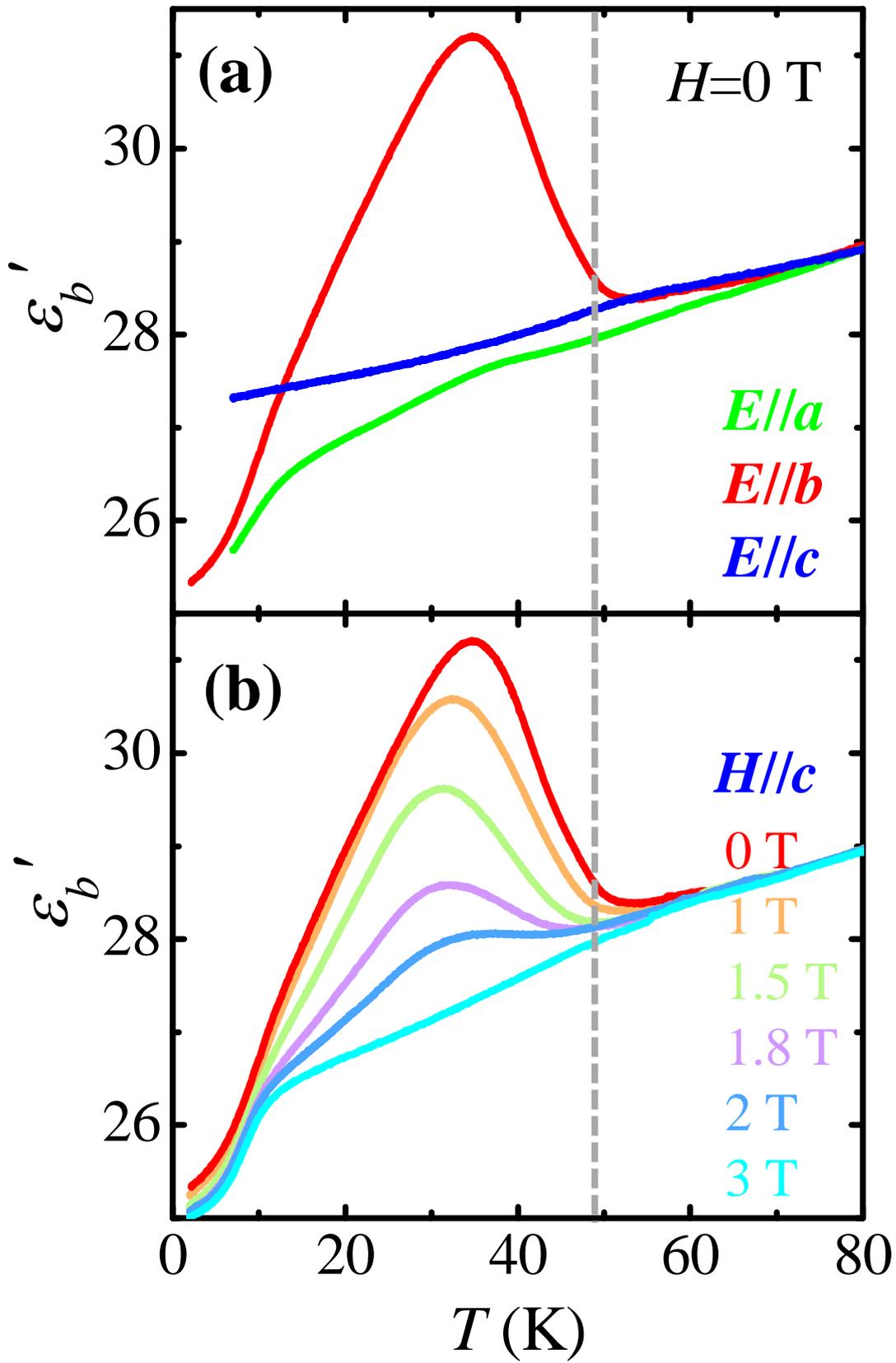

Fig. 3 N. Lee *et al.*

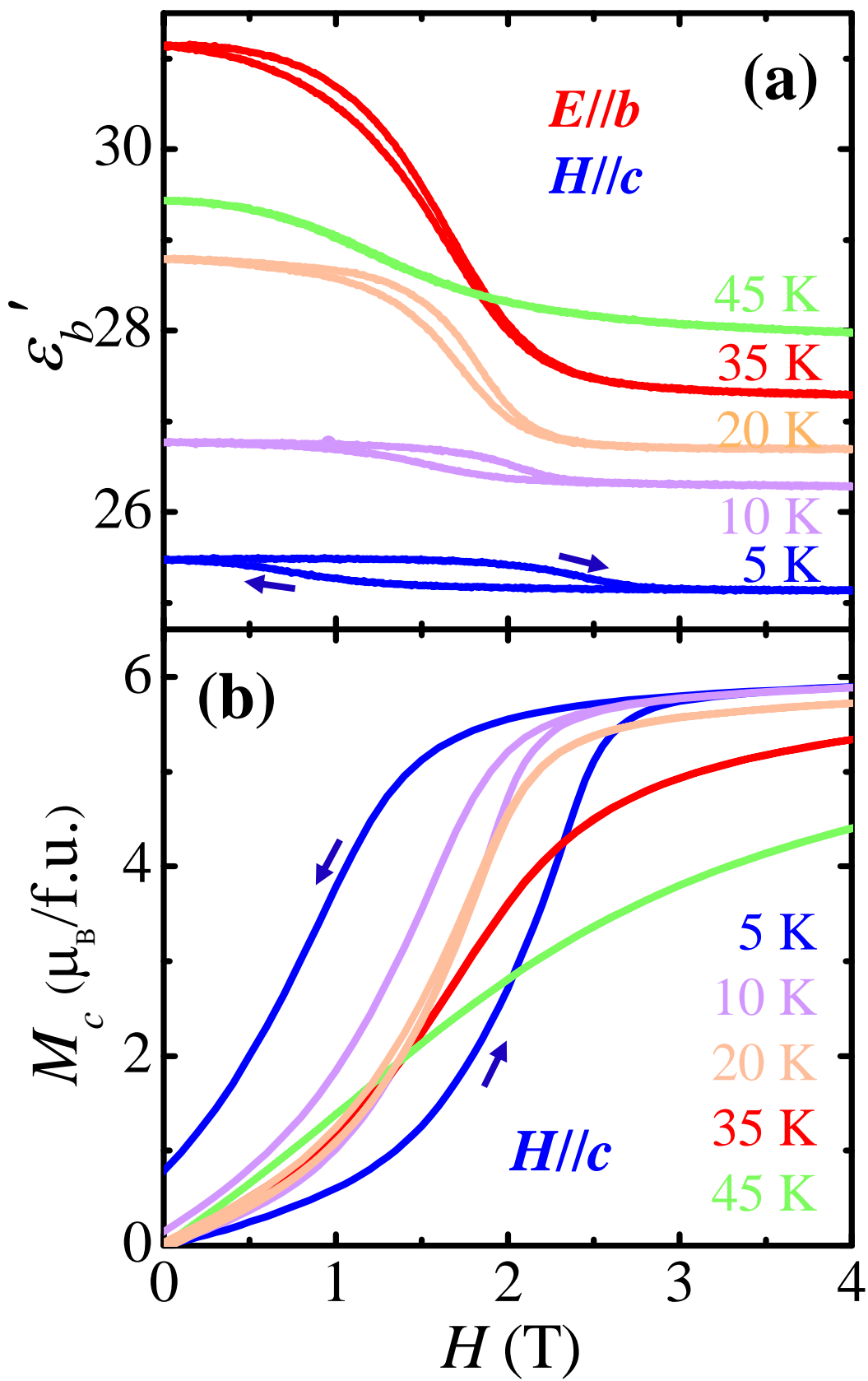

Fig. 4 N. Lee *et al.*